\documentclass[10pt]{an}
\usepackage{times}
\usepackage{natbib}
\usepackage{url}
\bibpunct{(}{)}{;}{a}{}{,}
\usepackage{graphicx}

\newcommand\muhz{\ensuremath{\mu\mathrm{Hz}}}
\newcommand\chisq{\ensuremath{\chi^2}}
\newcommand{\hdf}{HD~49933}
\newcommand{\tauhiz}{\ensuremath{\tau_\mathrm{\scriptscriptstyle HIZ}}}
\newcommand{\taubcz}{\ensuremath{\tau_\mathrm{\scriptscriptstyle BCZ}}}
\newcommand{\phihiz}{\ensuremath{\phi_\mathrm{\scriptscriptstyle HIZ}}}
\newcommand{\phibcz}{\ensuremath{\phi_\mathrm{\scriptscriptstyle BCZ}}}
\newcommand{\ahiz}{\ensuremath{A_\mathrm{\scriptscriptstyle HIZ}}}
\newcommand{\abcz}{\ensuremath{A_\mathrm{\scriptscriptstyle BCZ}}}

\def\myfigure#1#2#3#4{
	\begin{figure#4}
	\resizebox{1.00\hsize}{!}{\includegraphics{#1}}
	\caption{#2 \label{#3}}
	\end{figure#4}
}

\def\myfiguretwo#1#2#3#4#5{
	\begin{figure#5}
	\resizebox{0.50\hsize}{!}{\includegraphics{#1}}
	\resizebox{0.50\hsize}{!}{\includegraphics{#2}}
	\caption{#3 \label{#4}}
	\end{figure#5}
}

\begin{document}

\title{Model-independent
determination of sharp features
inside a star from its oscillation frequencies
}

\author{ 
     A.~Mazumdar\inst{1}
\thanks{anwesh@tifr.res.in}
\and
     E.~Michel\inst{2}
}

\authorrunning{Mazumdar, A. \and Michel, E.}
\titlerunning{Acoustic depths of sharp features inside a star}

\institute{
Homi Bhabha Centre for Science Education, TIFR,
V.~N.~Purav Marg, Mankhurd, Mumbai 400088, India.
\and
LESIA, Observatoire de Paris, 5 Place Jules Janssen, 
Meudon Cedex, 92195, France.
}

\abstract
{ It has been established earlier that sharp features like the base of
the convective zone or the second helium ionisation zone inside a star
give rise to sinusoidal oscillations in the frequencies of pulsation.
The acoustic depth of such features can be estimated from this
oscillatory signal in the frequencies. We apply this technique for the
CoRoT\thanks{The CoRoT space mission was developed and is operated by
the French space agency CNES, with the participation of ESA's RSSD and
Science Programmes, Austria, Belgium, Brazil, Germany, and Spain.}
frequencies of the solar-type star \hdf. This is the first time that
such analysis has been done of seismic data for any star other than the
Sun.  We are able to determine the acoustic depth of both the base of
the convective zone and the HeII ionisation zone of \hdf\ within $10\%$
error from the second differences of the frequencies. The locations of
these layers using this technique is in agreement with the current
seismic models of \hdf.  }


\maketitle

\section{Introduction}
\label{sec:intro}

A discontinuity in the derivatives of the sound speed in the stellar
interior introduces a characteristic oscillatory signature in the
frequencies of low degree modes of the form \citep{gough90}
\[ \sin (4\pi\nu\tau + \phi) \]
where $\tau$ is the acoustic depth of the layer, lying at a radius 
$r$, given by 
\[
\tau = \int_r^R \mathrm{d}r/c
\]
and $\nu$ is the frequency of oscillation. $\phi$ is a phase factor. 

In solar-type stars, the base of the convective envelope and the HeII
ionisation zones are the most prominent layers of such discontinuity.
Each of these layers contributes to a sinusoidal variation in the
frequencies of pulsation with the periodicity corresponding to the
respective acoustic depth of the layer.  The resulting combined signal
in the frequencies can be fitted to a suitable functional form to
extract the acoustic depths of these layers separately.  Here we apply
this technique to estimate the acoustic depths of these regions of the
solar-type star \hdf\ from its frequencies obtained from the CoRoT
satellite \citep{baglin06}. Although this method has been used with
great success for the Sun \citep[see e.g.,][]{ban94,mct94}, this is the
first instance of an application to a distant star. This was one of the
main objectives of the CoRoT seismology programme \citep{michel06}.

The significance of this method is that it does not depend on
theoretical stellar models -- the position of the sharp features are
estimated directly from the frequencies themselves. In fact, such
independent determination of these layers can help in the modelling of
the star, as illustrated by \citet{mazumdar05}. We describe the
technique used in this work and the results in the next two sections.

\section{Technique}
\label{sec:technique}

This oscillatory signal in the frequencies is quite small and is
embedded in the frequencies together with a smooth variation of the
frequencies arising from the regular variation of the sound speed in the
stellar interior.  It can be enhanced by using the second differences
\citep[see, e.g.,][]{ban94,ma01,basu04},
\begin{equation}
\delta^2 \nu(n,\ell) = \nu(n-1,\ell) - 2\nu(n,\ell) + \nu(n+1,\ell),
\end{equation}
instead of the frequencies themselves. Using the second differences
serves the dual purpose of removing the smooth component in frequencies,
and magnifying the oscillatory components. However, the errors in the
frequencies are also enhanced in taking the differences, which might
obscure the oscillatory signal.  This is the reason why one cannot use
even higher differences.

The acoustic depths of the base of the convective zone (BCZ) and the
second helium ionisation zone (HIZ), \taubcz\ and \tauhiz, respectively,
can be obtained from the data by fitting the second differences to a
suitable function representing the oscillatory signals from these layers
\citep{ma01}. We choose the following function along the lines of
\citet{basu04}: 
\begin{eqnarray}
\delta^2 \nu & = & a_0 
+ b_0 \sin (4\pi\nu\taubcz\!+\!\phibcz) \nonumber \\
& & + (c_0\!+\!c_1/\nu) \sin(4\pi\nu\tauhiz\!+\!\phihiz)
\label{eq:func}
\end{eqnarray}

This function has three components corresponding to a residual smooth
variation, and the two oscillatory components corresponding to BCZ and
HIZ. We do not use all the parameters in the function suggested by
\citet{basu04} to accurately represent the second differences because
one cannot afford to use too many free parameters for a relatively small
data set. The number of free parameters is optimised to strike a balance
between a fair representation of the oscillatory signal and a reasonable
\chisq. We find that the smooth component is fairly constant over the
range of frequencies that we use, as is the amplitude of the signal from
the BCZ.  However, the amplitude of the signal from the HIZ varies more
sharply with frequency, and thus requires at least one
frequency-dependent term. We note that \citet{basu04} have shown that
the exact form of the amplitudes of the oscillatory signal does not
affect the results significantly. 

The fit is carried out through a nonlinear \chisq\ minimisation,
weighted by the errors in the data. The errors in the second differences
are correlated, and this is taken care of by defining the \chisq\ using
a covariance matrix.  The contribution of the errors are considered by
producing $1000$ realisations of the data, where the mean values of the
frequencies are skewed by random errors corresponding to a normal
distribution.  The successful convergence of such a non-linear fitting
procedure is somewhat dependent on the choice of reasonable initial
guesses. To remove the effect of initial guesses affecting the final
fitted parameters, we carry out the fit for multiple combinations of
starting values.  For each realisation, $100$ combinations of initial
guesses are tried for fitting the function above and the one with the
minimum \chisq\ is accepted as the fit for that particular realisation.

The median value of each parameter for its distribution over $1000$
realisations is taken as the average. The error in the parameter is
estimated from the range of values covering $68\%$ area about the median
(corresponding to $1\sigma$ error). As an example, the histogram of
distribution of the parameters \taubcz\ and \tauhiz\ over $1000$
realisations are illustrated in Figs.~\ref{fig:histogram_benomar} and
\ref{fig:gruber}. The quoted errors in these parameters reflect the
width of these histograms on two sides of the median value.

\section{Results}
\label{sec:results}

Our primary data set is that of \citet{benomar09}. To check for
consistency we also use the data set of \citet{gruber09}, and the
theoretical frequencies of a model of \hdf\ (\citealt{deheuvels10}; see
also \citealt{goupil10}).  

\myfigure{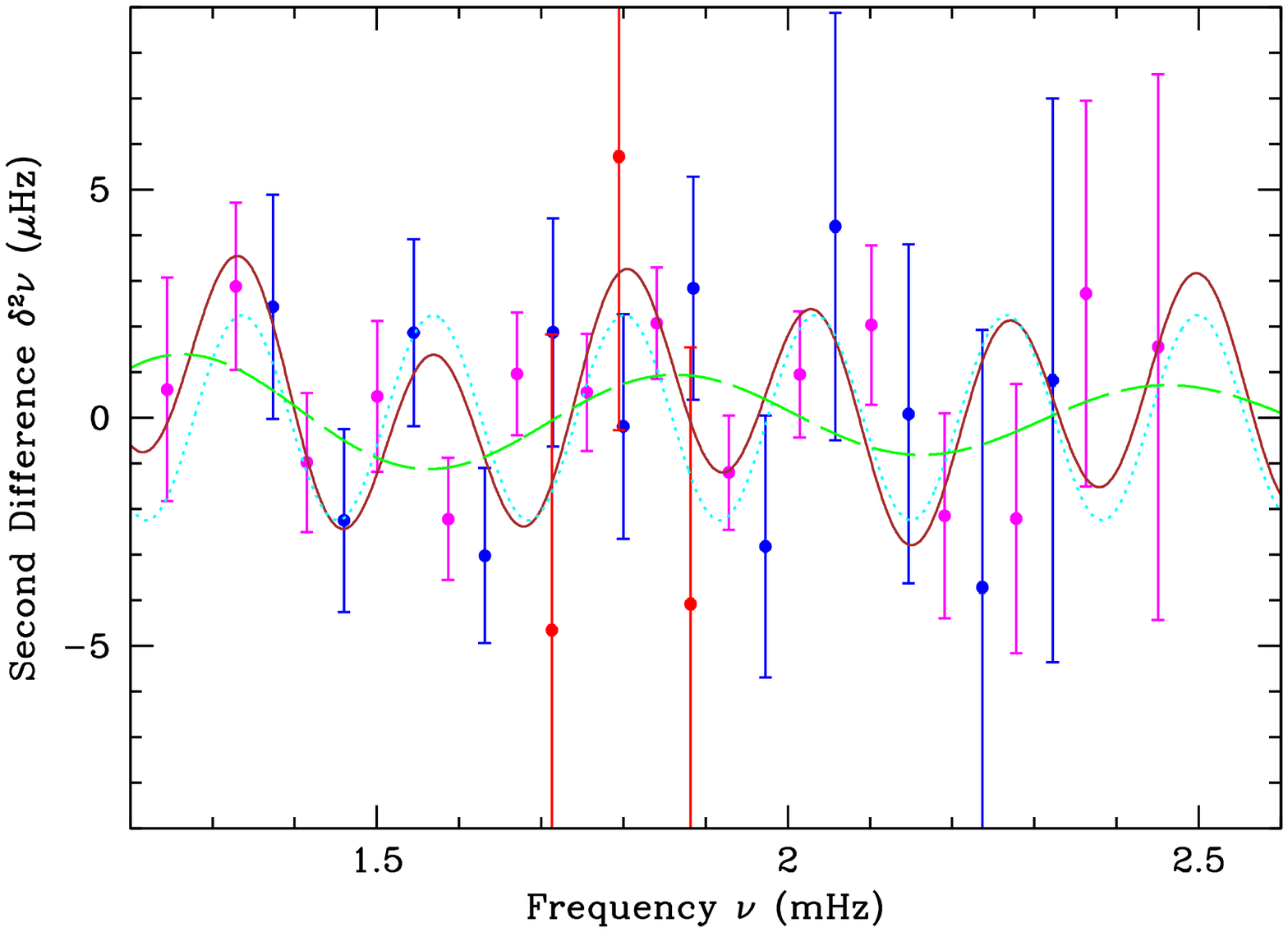}
{Second differences of frequencies of \hdf\ from $180$
days of CoRoT data \citep{benomar09} are plotted against frequencies
for $\ell=0$ ({\it blue} points), $\ell=1$ ({\it magenta} points), 
and a few $\ell=2$ ({\it red} points) modes.
The fit of the oscillatory function is shown by the {\it brown solid} line. 
The {\it cyan dotted} line shows the fitted oscillatory signal due to the 
base of the convection zone (BCZ) and the {\it green dashed} line shows 
the signal due to the second helium ionisation zone (HIZ)}
{fig:fit_benomar}
{}

\myfigure{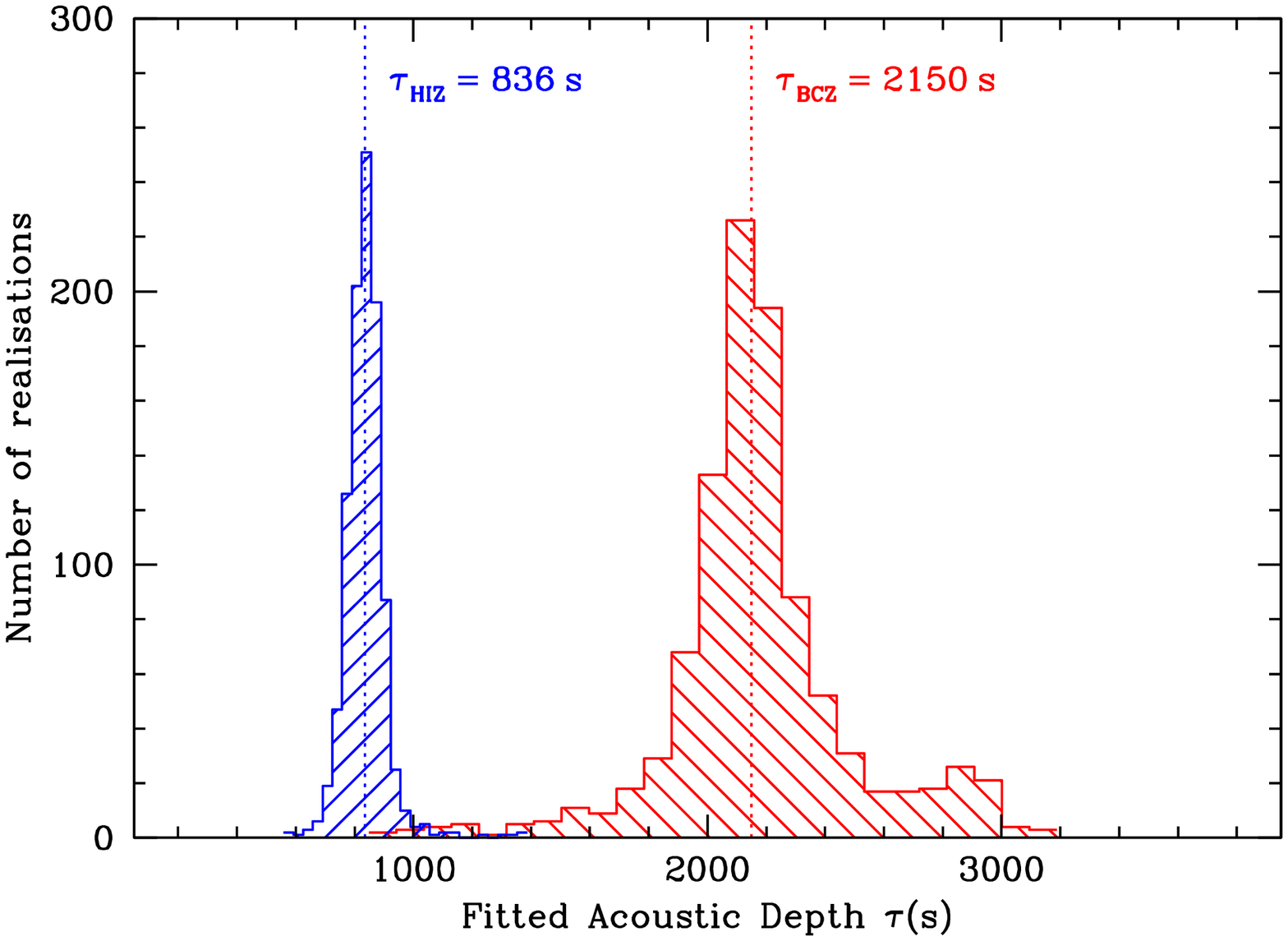}
{Histograms of fitted values of acoustic depths, \taubcz\ ({\it red}) 
and \tauhiz\ ({\it blue}), obtained for $1000$ different realisations
of the data according to \citet{benomar09} are shown. 
The median values are shown with the dotted lines. }
{fig:histogram_benomar}
{}

\myfiguretwo{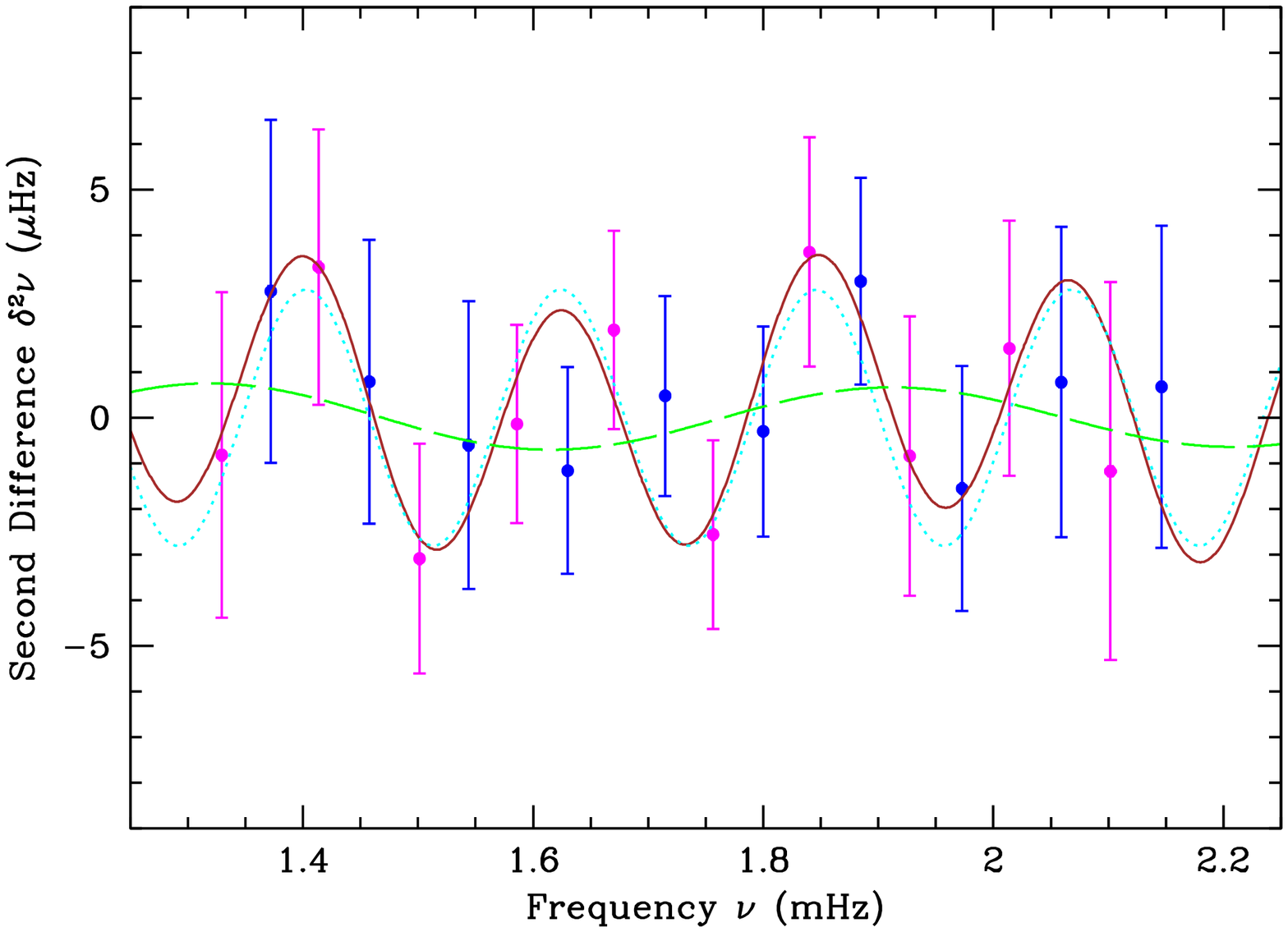}{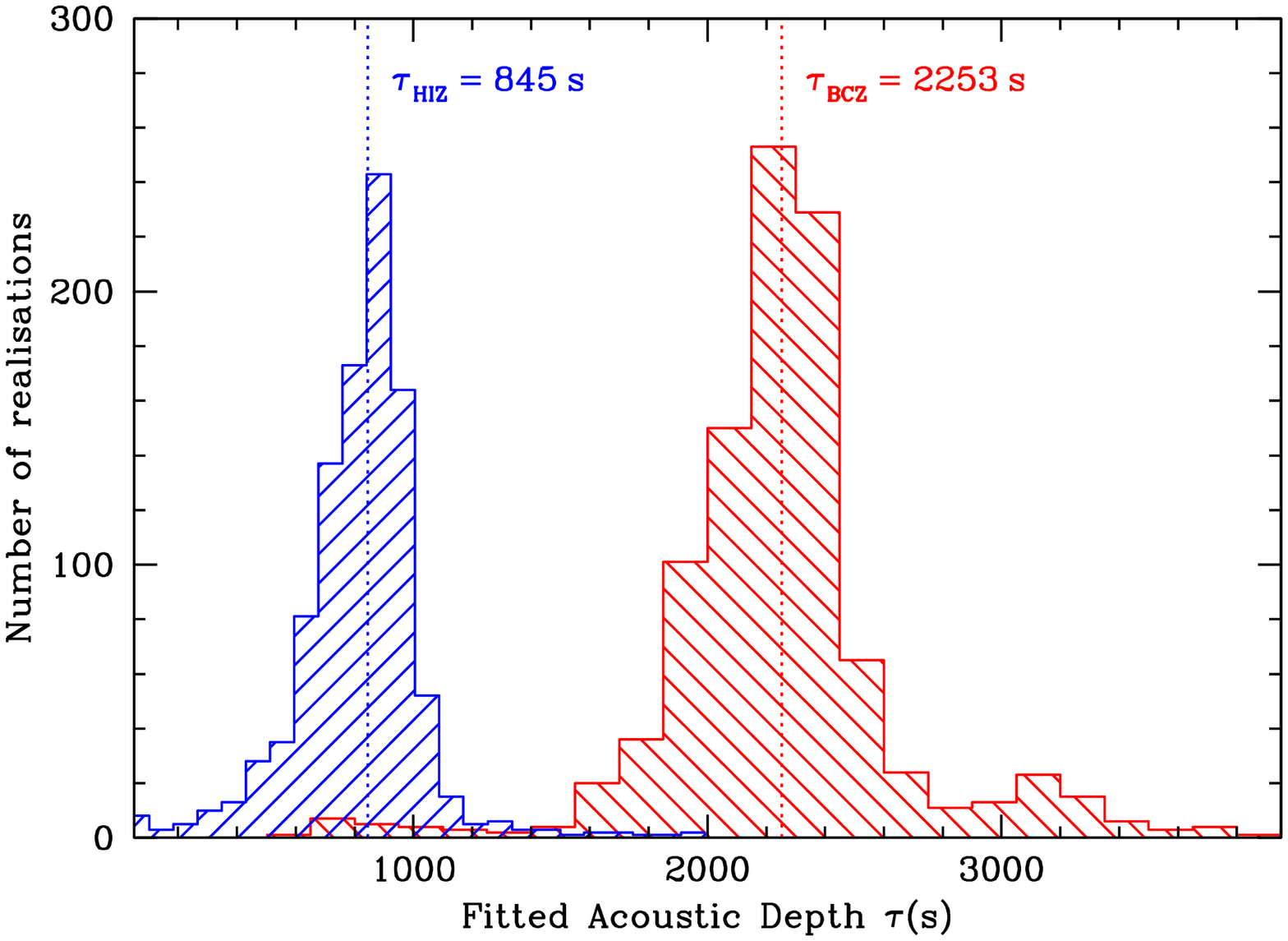}
{Similar plots as Figs.~\ref{fig:fit_benomar} and
\ref{fig:histogram_benomar}, but with the frequencies taken from the
60 day CoRoT data \citep{gruber09}. }
{fig:gruber}
{*}

The fit of the function given in Eq.~\ref{eq:func} is shown in
Fig.~\ref{fig:fit_benomar} for the data set from \citet{benomar09}. This
set consists of $31$ second differences computed from the individual
frequencies. We did not use frequencies which had errors of more than
$3\,\muhz$. The data represented in this figure corresponds to the mean
frequencies. In reality, different realisations of the data are fitted
independently. The figure also shows the function given in
Eq.~\ref{eq:func}, with the parameter values set equal to the median
values obtained from $1000$ realisations. The histograms of \taubcz\ and
\tauhiz\ values over all realisations are shown in
Fig.~\ref{fig:histogram_benomar}. The median values are indicated, and
the errors are estimated by considering the extent of the histogram
around the median such that $34\%$ of the cases are covered on each
side. Since the histograms are asymmetric, naturally this yields unequal
error bars. We consider these to be representative of $1\sigma$ error on
the median value of the parameter concerned. The average amplitude of
the oscillatory components corresponding to the two regions, \abcz\ and
\ahiz, are evaluated in terms of the frequencies themselves, instead of
the second differences \citep[see][for details]{ma01}. These results are
summarised in Table~\ref{tab:results}.

To check for consistency, we also used the data set from
\citet{gruber09}, which yielded $20$ second differences.  This data set
yields very similar results, shown in Fig.~\ref{fig:gruber} and listed
in Table~\ref{tab:results}, although with higher error bars, showing
that the method works even with lesser data.

Lastly, we also fit Eq.~\ref{eq:func} to the theoretical frequencies
computed from a stellar model of \hdf\ \citep{deheuvels10,goupil10}.
The actual model values of the acoustic depths, \taubcz\ and \tauhiz,
are remarkably close to the values obtained from this fit (see
Table~\ref{tab:results}). This is not entirely unexpected, since the
stellar model is obtained by matching the seismic frequencies
themselves. However, the methods used in the seismic modelling do not
explicitly involve determining the depths of the base of the convective
envelope or the second helium ionisation region. In contrast, our method
focuses directly on these two layers, irrespective of the stratification
in the rest of the star.  However, as pointed out by \citet{mazumdar05},
the location of these layers cannot be independent of the structure of
the rest of the star, and is thus linked implicitly to the detailed
model. But the process of fitting the acoustic depths does not involve
stellar modelling and thus the results obtained in this method are
completely model-independent.

\myfigure{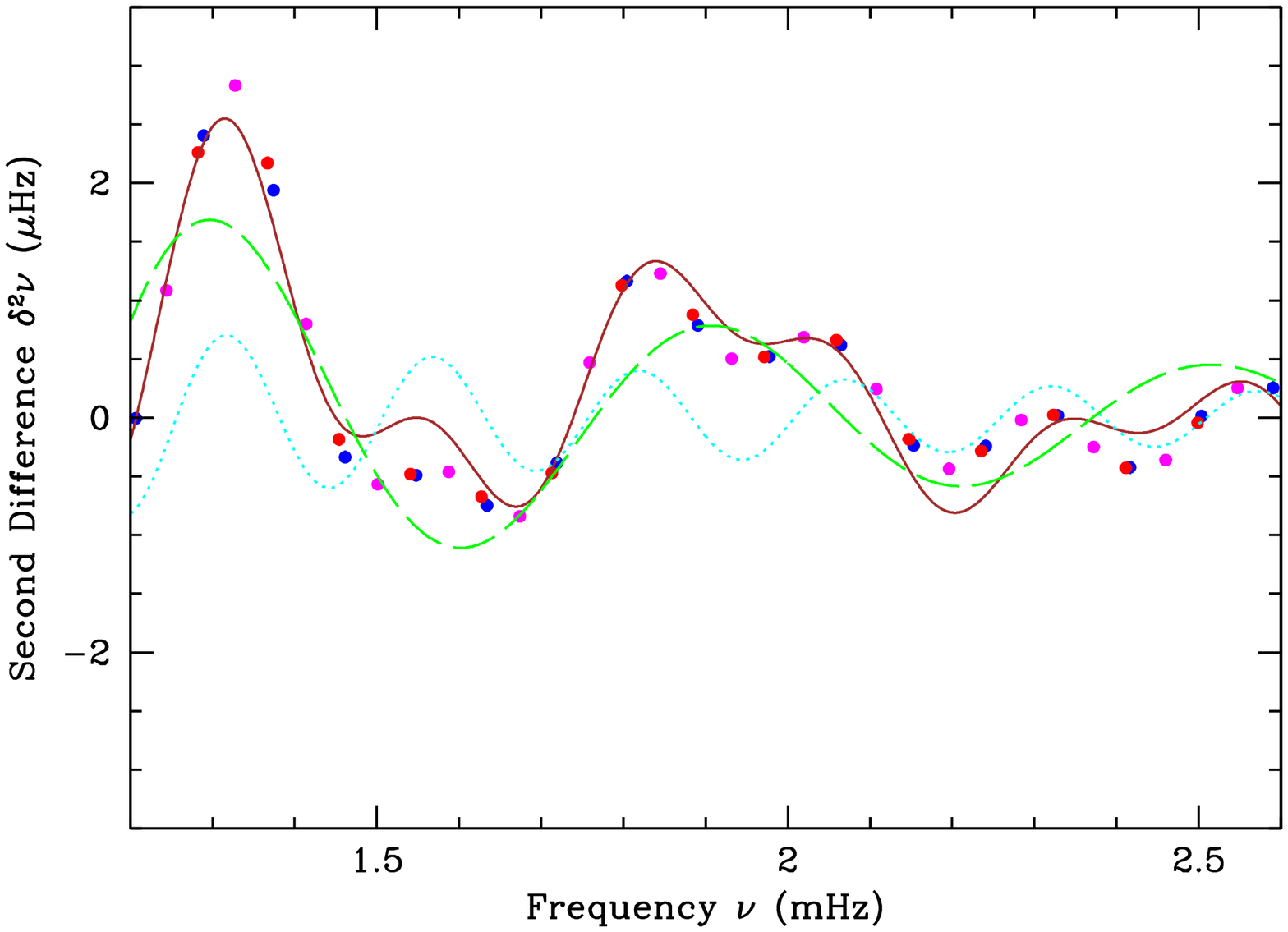}
{Second differences of frequencies of a theoretical model of
\hdf\ \citep{deheuvels10,goupil10} are plotted against frequencies for $\ell=0$,
$\ell=1$, and $\ell=2$ modes. The composite fit 
of the the oscillatory function, as well as its two components are
shown, similar to Fig.~\ref{fig:fit_benomar}. }
{fig:fit_model}
{}

\setlength{\tabcolsep}{0.45em}
\begin{table*}
\caption{Results of the fitting of second differences to
Eq.~\ref{eq:func} for two data sets and a theoretical model. The
quantities in parentheses represent the errors
in each parameter for the first two rows, and the actual theoretical model 
values in the third row.
\label{tab:results}
}
\begin{tabular}{lcccccc}
\hline
Data Set & No. of &
\multicolumn{2}{c}{Acoustic Depth} & \multicolumn{2}{c}{Average
Amplitude in $\nu$} & \chisq\\
& data pts. & \taubcz\ (s) & \tauhiz\ (s) & \abcz (\muhz)
& \ahiz (\muhz) & \\
\hline
CoRoT $180$~d \citep{benomar09} & $31$ & $2150$ ($+182$/$-257$) & $836$
($+56$/$-52$) & $0.57$ ($\pm 0.30$) & $1.29$ ($\pm 0.30$) & $1.03$\\
CoRoT $60$~d \citep{gruber09} & $20$ & $2253$ ($+289$/$-226$) & $845$ 
($+191$/$-142$) & $0.52$ ($\pm 0.20$) & $0.93$ ($\pm 0.32$) & $0.83$\\
Model \citep{deheuvels10,goupil10} & $57$ & $1995$ ($1999$) & $826$
($883$) & $0.46$ & $1.35$\\
\hline
\end{tabular}
\end{table*}

\section{Summary}
\label{sec:summary}

We have determined the acoustic positions of the base of the convective
envelope and the second helium ionisation zone in the star \hdf\ from
the oscillatory signal in its frequencies observed with CoRoT. The
method involves fitting a suitable function to represent the two
oscillatory signals arising from the two zones to the second differences
of the frequencies. We find almost similar results for two data sets,
one corresponding to only the initial 60-day run of CoRoT and another a
180-day combined time-series obtained with a long run of CoRoT. This is
the first instance where such a technique has been successfully applied
to a star other than the Sun.

Although these results are independent of any stellar modelling, they
match well with a seismic model of the star. In fact the acoustic depths
of such layers of sharp changes in the sound speed inside the star
extracted in this method can help to model the star \citep{mazumdar05}.

The same technique can be used to estimate the acoustic depths of such
layers in other stars. We have established that a time series of
moderate length, as obtained by CoRoT, is sufficient for the successful
application of this technique. A longer time series, as expected from
the Kepler mission, for example, would further reduce the errors on the
extracted acoustic depths.

\acknowledgements
This work was supported by the National Initiative on Undergraduate
Science (NIUS) undertaken by the Homi Bhabha Centre for Science
Education -- Tata Institute of Fundamental Research (HBCSE-TIFR),
Mumbai, India. We acknowledge support from Centre National d'Etudes
Spatiales (CNES).  AM acknowledges support from LESIA, CNRS during
visits to Observatoire de Paris, Meudon.

\end{document}